\documentclass[12pt,twoside]{article}   
\usepackage[super,sort,comma]{natbib}
\usepackage{lscape}
\usepackage{booktabs}
\usepackage{amsmath,amssymb,amsfonts}
\usepackage{fancyhdr}		

\usepackage[section]{placeins}   %
\makeatletter \renewcommand\@biblabel[1]{$^{#1}$} \makeatother
\newcommand{\cen}[1]{\begin{center} #1 \end{center}}

\usepackage[mathlines]{lineno}

\usepackage{hyperref}
\hypersetup{ colorlinks,
    citecolor=blue,
    filecolor=blue,
    linkcolor=blue,
    urlcolor=blue
}

\usepackage{graphicx}
\usepackage{xcolor}

\definecolor{gray}{rgb}{0.6,0.6,0.6}
\definecolor{red}{rgb}{0.85,0,0}
\definecolor{green}{rgb}{0,0.85,0}
\definecolor{blue}{rgb}{0,0,0.85}
\definecolor{beige}{rgb}{0.92,0.87,0.78}
\usepackage[all]{hypcap}    

\begin{document}
\cen{\sf {\Large {\bfseries A knee cannot have lung disease: out-of-distribution detection with in-distribution voting using the medical example of chest X-ray classification } \\
\vspace*{10mm}
Alessandro Wollek\thanks{1}, Theresa Willem\thanks{2}, Michael Ingrisch\thanks{3}, Bastian Sabel\thanks{3}, Tobias Lasser\thanks{1}
} \\
\footnotemark[1]\footnotemark[5]{Munich Institute of Biomedical Engineering and the School of Computation, Information, and Technology, Technical University of Munich}
\footnotemark[2]{Institute for History and Ethics in Medicine and Munich School of Technology in Society, Technical University of Munich}
\footnotemark[3]\footnotemark[4]{Department of Radiology, University Hospital Ludwig-Maximilians-Universit\"at}
\vspace{5mm}\\
Version typeset \today\\
}

\pagenumbering{roman}
\setcounter{page}{1}
\pagestyle{plain}
Corresponding author: Alessandro Wollek email: alessandro.wollek@tum.de address: Boltzmannstr. 11, 85748 Garching, Germany

\begin{abstract}
\noindent {\bf Background:}   Deep learning models are being applied to more and more use cases with astonishing success stories, but how do they perform in the real world?
Models are typically tested on specific cleaned data sets, but when deployed in the real world, the model will encounter unexpected, out-of-distribution (OOD) data.
\\
{\bf Purpose:}
To investigate the impact of OOD radiographs on existing chest X-ray classification models and to increase their robustness against OOD data.
\\
{\bf Methods:}
The study employed the commonly used chest X-ray classification model, CheXnet, trained on the chest X-ray 14 data set, and tested its robustness against OOD data using three public radiography data sets: IRMA, Bone Age, and MURA, and the ImageNet data set.
To detect OOD data for multi-label classification, we proposed in-distribution voting (IDV).
The OOD detection performance is measured across data sets using the area under the receiver operating characteristic curve (AUC) analysis and compared with Mahalanobis-based OOD detection, MaxLogit, MaxEnergy and self-supervised OOD detection (SS OOD).
\\
{\bf Results:}
Without additional OOD detection, the chest X-ray classifier failed to discard any OOD images, with an AUC of 0.5.
The proposed IDV approach trained on ID (chest X-ray 14) and OOD data (IRMA and ImageNet) achieved, on average, 0.999 OOD AUC across the three data sets, surpassing all other OOD detection methods.
Mahalanobis-based OOD detection achieved an average OOD detection AUC of 0.982.
IDV trained solely with a few thousand ImageNet images had an AUC 0.913, which was higher than MaxLogit (0.726), MaxEnergy (0.724), and SS OOD (0.476).
\\
{\bf Conclusions:}
The performance of all tested OOD detection methods did not translate well to radiography data sets, except Mahalanobis-based OOD detection and the proposed IDV method.
Training solely on ID data led to incorrect classification of OOD images as ID, resulting in increased false positive rates.
IDV substantially improved the model's ID classification performance, even when trained with data that will not occur in the intended use case or test set, without additional inference overhead.
The corresponding code is available at \url{https://gitlab.lrz.de/IP/a-knee-cannot-have-lung-disease}.
\\
\end{abstract}
\tableofcontents

\newpage

\setlength{\baselineskip}{0.7cm}      

\pagenumbering{arabic}
\setcounter{page}{1}
\pagestyle{fancy}

\section{Introduction}\label{sec:introduction}
\begin{figure}[h]
\centering
\includegraphics[width=\textwidth]{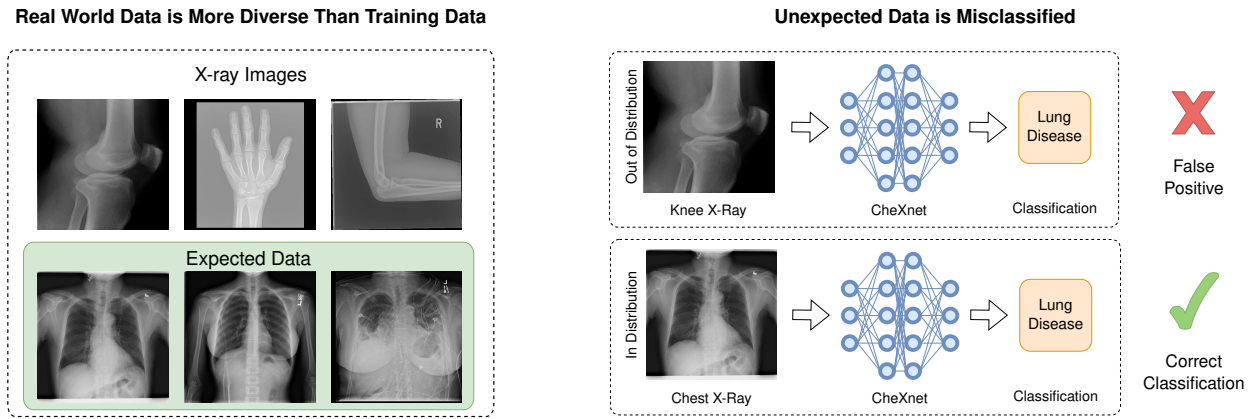}
\caption{
Deep learning models in the real world must be able to handle OOD data.
(\textbf{left}) A chest X-ray classifier (CheXnet) is trained on chest X-rays and tested with expected production data: chest X-rays.
In a clinic, the model has to handle non chest X-ray images confidently, as the data cannot be manually cleaned beforehand.
(\textbf{right}) A model trained and tested only on chest X-ray images will incorrectly classify OOD images (here: an X-ray of a knee) as having lung disease.
}\label{Figure0}
\end{figure}

Modern machine learning models are achieving great successes in real world medical applications, such as diabetic retinopathy diagnosis~\cite{gulshanDevelopmentValidationDeep2016}, skin cancer classification~\cite{ estevaDermatologistlevelClassificationSkin2017}, or lung disease assessment~\cite{rajpurkarChexnetRadiologistlevelPneumonia2017, rajpurkarDeepLearningChest2018, majkowskaChestRadiographInterpretation2019}.
Due to the early and profound digitization of imaging techniques, machine learning in radiology can already show convincing successes, such as the detection of certain critical pathologies of the lung on X-ray images with performance non-inferior to radiologists~\cite{rajpurkarChexnetRadiologistlevelPneumonia2017}.
Considering the increasing demand for imaging, while the number of radiologists remains insufficient, the aforementioned and similar models can help improve medical patient care, for example, by screening acquired radiographs for critical findings prior to radiologist interpretation~\cite{aliDiagnosticRadiologyLiberia2015, idowuDiagnosticRadiologyNigeria2020, rosmanImagingLand10002015, rosenkrantzUSRadiologistWorkforce2016, rimmerRadiologistShortageLeaves2017, wollekAttentionbasedSaliencyMaps2023}.
Then, patients with time sensitive illnesses will receive treatment earlier, potentially saving their lives.

What all of these chest X-ray classifiers have seen, once trained, validated and tested, are chest X-rays of a certain type, the in-distribution (ID) images.
Consequently, the features learned depend on the assumption that the input is ID.
But despite the advanced level of digitization, individual workflows for creating and archiving radiological images and linking them to other patient data are subject to manual intervention by staff and are consequently prone to human error, breaking this assumption.
Just one example would be that mixed-up labeling can arise of patients for whom X-ray images of several body parts have been taken.
Consequently, images of a knee joint, for example, would be fed to a model for detecting pulmonary pathologies.
Hence, in the aforementioned scenario, the presentation of out-of-distribution (OOD) images, erroneous and potentially patient-harming events are possible.

A major problem of current deep learning models is that they make high confidence predictions when facing unexpected (OOD) data, like a knee X-ray~\cite{nguyenDeepNeuralNetworks2015a, nalisnickDeepGenerativeModels2018, hendrycksNaturalAdversarialExamples2021}.
In our scenario, prioritization based on false, high-confidence, OOD X-rays can lead to longer waiting times for other patients with time critical conditions, like a pneumothorax, potentially risking their life until the error is discovered and resolved.
Moreover, repeated instances of such misreporting will - if not balanced with transparency measures sufficiently, e.g., using saliency maps \cite{wollekAttentionbasedSaliencyMaps2023} -\ quickly lead physicians to distrust the model, eventually leading them to refrain from using it~\cite{robinetteEffectRobotPerformance2017, vayenaMachineLearningMedicine2018, novTransformationPatientclinicianRelationships2021}.

Therefore, in recent years, several methods, have been proposed to detect OOD samples~\cite{hendrycksBaselineDetectingMisclassified2017, hendrycksScalingOutofDistributionDetection2020, wangCanMultilabelClassification2021, hendrycksDeepAnomalyDetection2019,hendrycksUsingSelfSupervisedLearning2019,calliFRODOIndepthAnalysis2022}.
Commonly, the OOD detector converts the output of a model to an ID score.
For example, one of the earliest approaches, Max. Probability~\cite{hendrycksScalingOutofDistributionDetection2020} uses the highest class probability as ID probability.
In their experiments, the authors noticed lower confidence scores for the highest class probability for OOD inputs compared to ID inputs.
Another approach, proposed by Lee et al.~\cite{leeSimpleUnifiedFramework2018}, also applied to chest X-rays~\cite{calliFRODOIndepthAnalysis2022}, models OOD data based on the smallest Mahalanobis distance between the input and a class conditional Gaussian distribution in the latent space.
Furthermore, Hendrycks et al. propose to use a self-supervised training scheme to improve OOD detection performance~\cite{hendrycksUsingSelfSupervisedLearning2019}.

So far, the problem caused by OOD data has been investigated mostly on toy data sets, for example, a model trained on the CIFAR-10 data set~\cite{krizhevskyLearningMultipleLayers2009}, learning to classify automobiles and trucks, is tested on the SVHN data set \cite{netzerReadingDigitsNatural2011} containing house numbers, or in-house data sets~\cite{calliFRODOIndepthAnalysis2022}.
This raises the question if the test performance of proposed OOD detectors translate to an existing model trained on chest X-rays.
Figure~\ref{Figure0} motivates this problem: as the real world data consists of more than frontal chest X-rays, a classifier like CheXnet~\cite{rajpurkarChexnetRadiologistlevelPneumonia2017} must handle OOD images safely.
\c{C}all\i{} et al.\ investigated the effect of an in-house collected OOD X-ray data set on the task of nodule classification, localization and lung segmentation~\cite{calliFRODOIndepthAnalysis2022}.
In contrast, we focus on the more general multi-label chest X-ray classification problem.

In this work, we are addressing the practical consequences of OOD data by examining the impact of non chest radiographs on the chest X-ray classifier CheXnet.
We selected this model as it performs similarly to radiologists~\cite{rajpurkarChexnetRadiologistlevelPneumonia2017,rajpurkarDeepLearningChest2018} and is widely used as a benchmark.
The major contributions of our work are: we systematically explore the OOD detection performance of the CheXnet chest X-ray classifier on three realistic OOD data sets; we show that the benchmark performance of current OOD detection methods mostly do not translate to this domain; and we demonstrate that our proposed method in-distribution voting (IDV) improves OOD detection and generalizes to other data sets.
\section{Methods}
\begin{figure}[t]
\centering
\includegraphics[width=\textwidth]{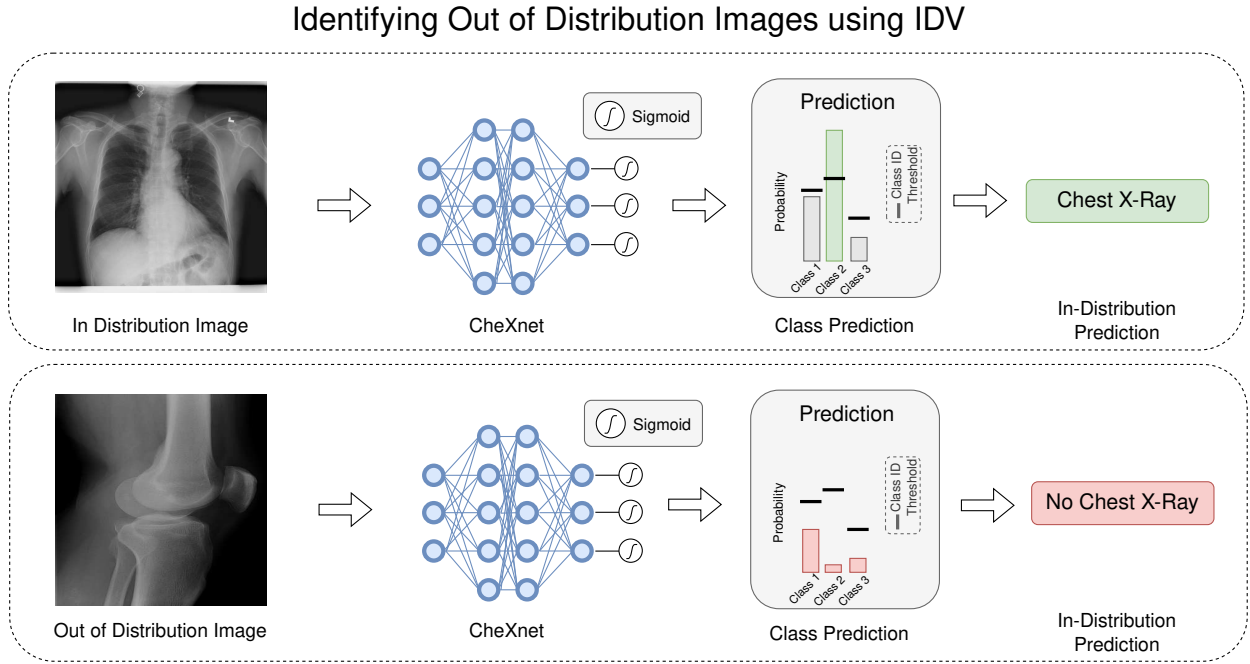}
\caption{
Out-of-distribution (OOD) detection using our proposed method in-distribution voting (IDV).
The model is trained with in-distribution (ID) and OOD images.
Before inference, class-wise ID threshold are used to classify the input images as either ID (chest X-ray, \textbf{top}) or OOD (knee, \textbf{bottom})
In this multi-label setting, a sample is classified as OOD only if all classes unanimously vote against ID.
}\label{FigureMethod}
\end{figure}

\subsection{Chest X-Ray Classification: CheXnet}\label{methods:chexnet}
Following \cite{rajpurkarChexnetRadiologistlevelPneumonia2017}, we fine-tuned a DenseNet-121~\cite{huangDenselyConnectedConvolutional2017a} on the CXR14 data set.
The model was pre-trained on ImageNet and the weights are available on \url{pytorch.org}.
For fine-tuning, we replaced the last layer with a fully-connected layer with 15 outputs, matching the 14 classes of the CXR14 data set plus the  ``no finding'' class which represents the absence of any of the 14 pathologies.
We expanded the original CheXnet architecture from 14 to 15 outputs to differentiate between ID ``no finding'' CXRs and OOD samples.
Since an image may exhibit signs of multiple pathologies, we modeled the classification task as a multi-label classification task, where each class is predicted independently.
The output scores were converted to a probability for each class by applying the sigmoid function: \[\sigma(x) = \frac{1}{1 + e^{-x}.}\]

We used binary cross entropy as loss function and trained the model using ADAM \cite{kingmaAdamMethodStochastic2015} optimization with default parameters ($\beta_1 = 0.9,\quad\beta_2 = 0.999$) and an initial learning rate of 0.0003.
We divided the learning rate by a factor of ten if the validation loss did not improve over the last two epochs.
We applied weight decay with a value of 0.0001.
We trained the model for eight epochs and selected the best model based on the mean area under the receiver operating characteristic curve (AUC) for classifying the ID validation data set.
The input images were resized to 256 $\times$ 256 pixels and normalized according to the ImageNet mean and standard deviation.
Then, we applied 224 $\times$ 224 ten crop, i.e., we took crops from each corner and the center of the image and repeated the process for the horizontally flipped image: producing ten 224 $\times$ 224 pixel images per sample.
The model predictions of the ten images were averaged before calculating the loss.

When including OOD images into the training data, the model must predict the absence of any pathology.
This is in contrast to healthy CXR images, where the ``no finding'' class must be predicted.
In the default CheXnet setup this would result in predicting the same for both OOD and CXR with no finding. Like the ID images, the OOD images are normalized according to the ImageNet mean and standard deviation and passed to the model in the same fashion as the ID images.

\subsection{Proposed Method: In-Distribution Voting}

To improve the robustness of CheXnet's predictions and OOD detection performance we propose in-distribution voting (\textbf{IDV}).
We classified a sample as ID if at least one class-wise prediction exceeds the class-wise ID threshold, as illustrated in Figure~\ref{FigureMethod}.

Unlike other multi-label OOD detection techniques in the literature, this approach leverages OOD data to separate actual image classification from OOD detection.
We adapted approaches proposed in the literature~\cite{hendrycksDeepAnomalyDetection2019,  bevandicSimultaneousSemanticSegmentation2019} and included OOD data in the training data set, known as outlier exposure~\cite{hendrycksDeepAnomalyDetection2019} or negative data~\cite{torralbaUnbiasedLookDataset2011}.
We motivate the use of OOD data during training to break the ``closed world'' assumption.
In other words, we forced the model not to condition the predictions on the chest X-ray input assumption.
In our case, the model was required to predict the absence of any class for OOD samples, resulting in a zero vector.

It is noteworthy that although ``no finding'' samples do not have any labeled classes, we consider them as ID, as they are chest X-rays.
For such CXRs that exhibit no indications of the 14 classes, the model had to predict the ``no finding'' class.
In our experiments, we also used unrealistic OOD samples such as photos from ImageNet during training, since they are expected to exist when employing a pre-trained model.

In a production setting, the ID thresholds would be set independently for each class utilizing the validation set that contains both ID and OOD images instead of calculating the AUC to report the general performance.
When training with OOD images, both training and validation splits are expanded to include the OOD training/validation splits.
\begin{figure}[ht]
\centering
\includegraphics[width=\textwidth]{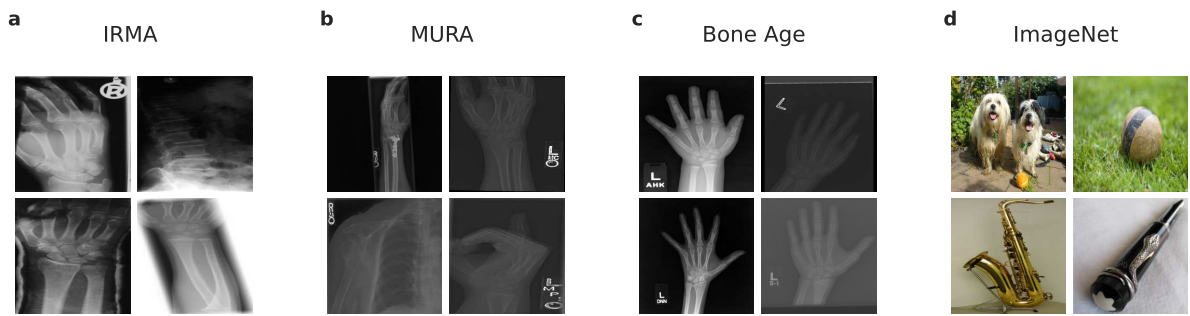}
\caption{
In this work, we utilized four out-of-distribution datasets: the Image Retrieval in Medical Applications (IRMA) data set~\cite{deserno15363IRMA2009} \textbf{(a)}, the Musculoskeletal Radiographs (MURA) data set~\cite{rajpurkarMURALargeDataset2018} \textbf{(b)}, the Bone Age data set~\cite{halabiRSNAPediatricBone2019} \textbf{(c)}, and the ImageNet data set~\cite{russakovskyImageNetLargeScale2015} \textbf{(d)}.
The IRMA data set comprises a diverse collection of radiographic images, while the MURA data set contains solely upper extremity radiographs, and the Bone Age data set consists of hand radiographs.
Lastly, the ImageNet data set is a collection of web-scraped photographs.
All four data sets are publicly available.
}\label{fig:datasets}
\end{figure}

\subsection{Other Out-of-Distribution Detection Methods}
In the literature, several methods for OOD detection have been proposed to explicitly filter OOD samples.
Hendrycks and Gimpel were among the first to tackle this problem~\cite{hendrycksBaselineDetectingMisclassified2017}.
They utilized the maximum value of the softmax prediction as the ID probability.
In their work, they justify this choice by observing that the highest prediction for OOD samples is lower than that of ID samples.
Since softmax is commonly used for single-label classification problems, they extended the approach to multi-label classification tasks by using the maximum logit of the classification layer (\textbf{Max. Logit})~\cite{hendrycksScalingOutofDistributionDetection2020}.
In contrast, Wang et al. used the label-wise energy function~\cite{liangEnhancingReliabilityOutofdistribution2018} instead of the sigmoid to transform the model output to an ID score (\textbf{Max. Energy})~\cite{wangCanMultilabelClassification2021}.

Instead of converting the model's output to an ID score, several approaches use the activations of the model to generate class-conditional Gaussian distributions (\textbf{Mahalanobis})~\cite{leeSimpleUnifiedFramework2018,calliFRODOIndepthAnalysis2022}.
This method models OOD images as unlikely points in the class distribution, i.e., having a large Mahalanobis distance to the modeled class means in the latent space.
Lee et al.~\cite{leeSimpleUnifiedFramework2018} motivated the use of the Mahalanobis distance between the mean representation of a class and the input in the feature space, instead of performing OOD detection in the label space due to ``label overfitting'', i.e., that the model predictions are conditioned on the training labels.
For Mahalanobis-based OOD detection, we use the output of the penultimate layer to determine the Mahalanobis scores similar to the work by \c{C}all\i{} et al~\cite{calliFRODOIndepthAnalysis2022}.

Hendrycks et al. propose training the classification model with additional self-supervised heads to improve OOD robustness (\textbf{SS OOD})~\cite{hendrycksUsingSelfSupervisedLearning2019}.
In this method, the model has to additionally predict image rotation and translation.
OOD detection is performed by taking the highest class prediction probability as ID score.

\begin{table}[h]
\centering
\begin{tabular}{@{}lr|rrrr|r@{}}
\hline
                & \multicolumn{1}{c}{In-} & \multicolumn{5}{c}{Out-of-Distribution} \\
Data Set        & CXR14 & IRMA  & MURA      & Bone Age  & ImageNet  & Subset  \\
\hline
Pre-processing  & -     & \multicolumn{1}{c}{Remove CXR}& -     & -         & Sample    & Sample \\
\hline
Training        & 78,468 & 3,088 & 35,366    & 8,179     & 217,818   & 3,088 \\
Validation      & 11,219 & 772   & 772       & 772       & 54,455    & 772 \\
Testing         & 22,433 & 3,860 & 3,860     & 3,860     & 3,860     & 3,860 \\
\hline
\textbf{Total}  & 112,120& 7,720 & 39,998    & 12,811    & 276,133   & 7,720 \\
\hline
\end{tabular}

\caption{
Data Sets used in our experiments.
The smallest out-of-distribution data set (IRMA) is split into 40/10/50 \% training/validation/testing.
To compare different scenarios we used the same number of images for validation and testing of the other OOD data sets (MURA~\cite{rajpurkarMURALargeDataset2018}, BoneAge~\cite{halabiRSNAPediatricBone2019}, ImageNet~\cite{russakovskyImageNetLargeScale2015}).
The remaining images were used for training.
Because the ImageNet data set is an order of magnitude larger than the ID CXR14 data set~\cite{wangChestXray8HospitalscaleChest2017} we took a random sample first.
To examine the different data set sizes we also fixed the size of the OOD training data splits to have the same amount of images (Subset). CXR = chest radiograph.
}\label{table:data set}
\end{table}

\subsection{Data Sets}\label{sec:datasets}
Not every OOD sample is equally likely in a real-world scenario.
In a production setting, the CheXnet model can encounter OOD X-ray images, as the distinction between ID and OOD X-ray images is based on manual, error-prone tagging.
Photographs on the other hand are not part of the image processing pipeline in a radiology department and can thus be assumed not to be found in a real-world scenario.

For our OOD detection experiments, we selected three publicly available radiographic data sets, IRMA~\cite{deserno15363IRMA2009}, MURA~\cite{rajpurkarMURALargeDataset2018}, and BoneAge~\cite{halabiRSNAPediatricBone2019}, containing X-ray images of various body parts as realistic OOD test data sets to test cross-data set generalization~\cite{torralbaUnbiasedLookDataset2011}.
We specifically chose publicly available data sets to ensure reproducability of our findings and encourage future work.
Further data set details are listed in Table~\ref{table:data set}.
\subsubsection{In-Distribution Chest X-ray 14}
We use the train-test split provided by the authors of the Chest X-ray 14 (CXR14) data set, having non-overlapping patients.
We further randomly split the provided training data set into training and validation sets, again with non-overlapping patients resulting in 78,468 training, 11,219 validation, and 22,433 test images (see also Table~\ref{table:data set}).
All three splits have a similar prevalence of class labels.
In summary, the original data set is split into 70 \% training, 10 \% validation, and 20 \% test data.
In contrast to the original CheXnet model, we expand the target classes and include ``no finding'' to differentiate between healthy CXR and other images.
We also use the images labeled as ``no finding'' for training, as 46 \% of the images are labeled as such.
For these images, the model must predict the absence of all 14 pathologies in the original CheXnet setup.

\subsubsection{Out-of-Distribution Data Sets}
For our OOD detection experiments we use the following data sets:

\begin{itemize}
    \item \textbf{IRMA}: the image retrieval in medical applications data set \cite{deserno15363IRMA2009} consists of 14,410 diverse radiographic images; 12,677 are annotated according to the anatomical category, 1733 are test images without annotation. The original task was to predict the correct anatomical category.
    \item \textbf{MURA}: the musculoskeletal radiographs data set \cite{rajpurkarMURALargeDataset2018} consists of 40,561 radiographic images, displaying different upper extremity bones. The original task was to predict if the X-ray study is normal or abnormal.
    \item \textbf{BoneAge}: the Bone Age data set \cite{halabiRSNAPediatricBone2019} consists of 12,811 hand radiographs of children.
        The original task was to predict the age of the patient.
  \item \textbf{ImageNet}: the ImageNet data set~\cite{russakovskyImageNetLargeScale2015} contains over one million web scraped photographs.
        The data set is often used for pre-training computer vision models.
        There are several tasks for this data set, including image classification and object detection.
\end{itemize}

While the CheXnet model has been pre-trained on predicting the ImageNet classes, they are OOD regarding the target task of chest X-ray classification, as the data set does not include chest X-rays.
Therefore, we use it as additional non chest X-ray OOD data set, allowing us to investigate the performance of our proposed method ``In-Distribution Voting'' (IDV).
This is relevant for use cases where no or only few realistic OOD images are available.
The OOD data sets are illustrated in Figure~\ref{fig:datasets}.

Because the IRMA data set is the smallest data set, we sample every OOD data set so that their test and validation split sizes match the IRMA splits. Furthermore, we create a subset of every OOD data set to account for training data size.
\paragraph{IRMA}
We only use the provided training images, as we require the IRMA labels to exclude ID chest radiographs from the data set. We remove all chest X-rays from the data set according to their anatomical code and exclude images with an anatomical code starting with 57, 75, 05, or 150, resulting in 7,720 images.
We split the remaining images randomly into training, validation, and testing using a 30 \% / 20 \% / 50 \% split to ensure enough images in the test split.
\paragraph{Bone Age}
We randomly sample the test and validation images according to the data split sizes of the IRMA data set (772 validation images, 3,860 test images, see Table~\ref{table:data set}). The remaining 8,179 images are either used all or sampled according to the IRMA training set size (3,088 images) for the training split.
\paragraph{MURA}
We split the MURA data set, containing 40,561 images, similar to the Bone Age data set: the validation and test partitions are randomly sampled, matching the size of the IRMA validation/test splits, listed in Table~\ref{table:data set}. Either all remaining images or ones sampled according to the IRMA training set size (3,088 images) are used for training.
\paragraph{ImageNet}
Due to the size of the ImageNet Large Scale Visual Recognition Challenge (ILSVRC) data set compared to the Chest X-ray 14 data set we use only 50 \% of the 544,546 images provided in the “LOC\_train\_solution.csv” file for training and another 20 \% for validation, see Table~\ref{table:data set}.
When accounting for OOD data set sizes, we sample the training and validation sets according to the size of the IRMA splits.
For both cases we sample the test set according to the IRMA test split size.
All train/validation/test splits were created with non-overlapping images.

\section{Results}
\begin{figure}[h]
\centering
\includegraphics[width=\textwidth]{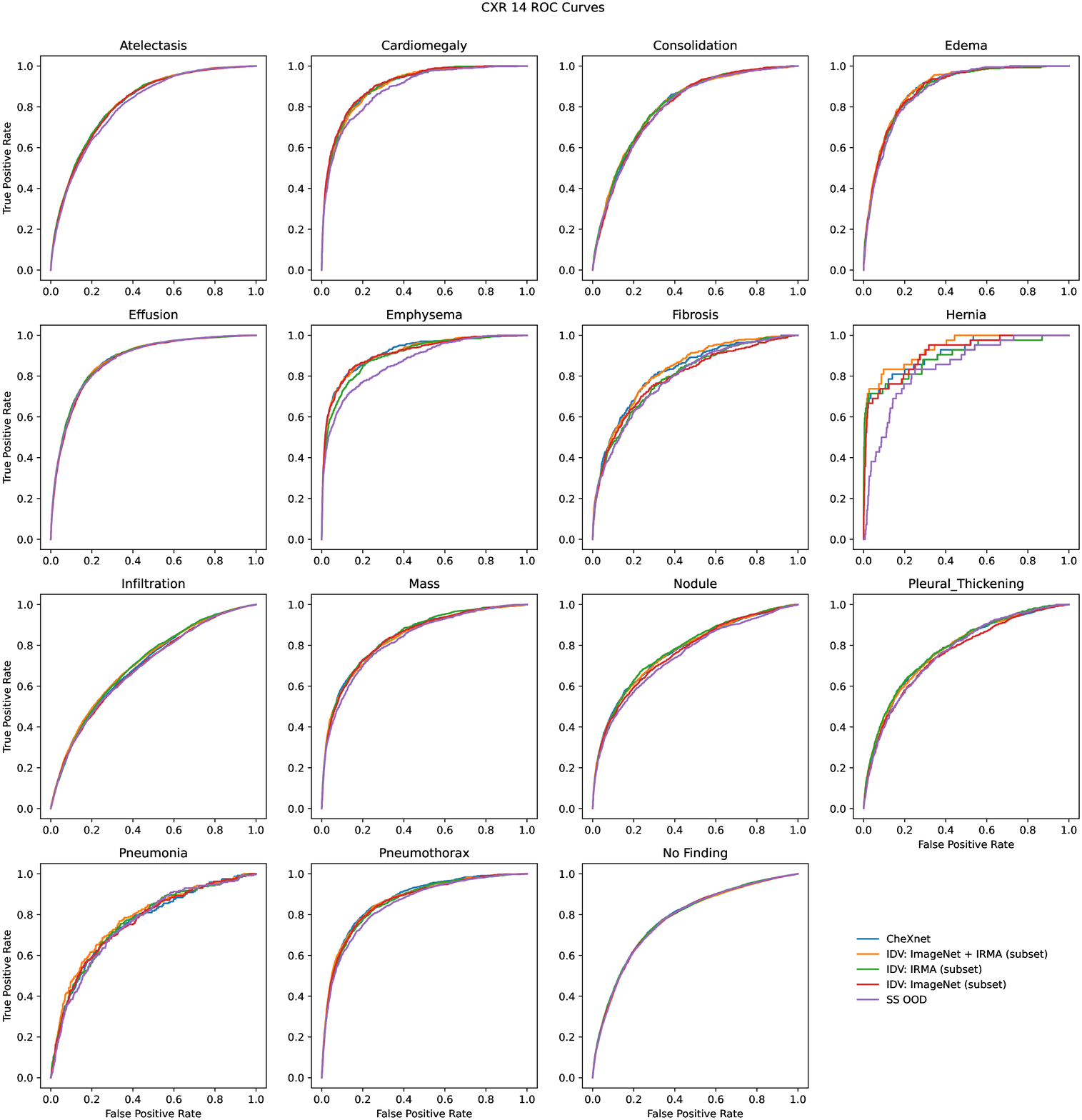}
\caption{
  ROC curves of all CXR14 classes for the main experiment settings.
  Training with OOD data, as proposed by our method, IDV, had no clear negative effect on CXR classification.
  In contrast, training with self-supervised heads (SS OOD) affected the classification negatively.
  The depicted IDV runs were trained with 3088 OOD images (subset).
  \\
  ROC = Receiver Operating Characteristic,
  OOD = out-of-distribution,
  IDV = in-distribution voting,
  CXR = Chest X-ray.
}\label{fig:roc_cxr14}
\end{figure}

\subsection{Chest X-ray Classification}

\begin{landscape}
\begin{table*}
\centering
\begin{tabular}{lrrrrrrrrrrrrr}
\toprule

      &  IDV &  IDV &IDV   &IDV   & CheXnet    &IDV   &IDV   & IDV        &IDV   & SS OOD   &IDV   &IDV   &IDV  \\
\midrule
 {Training Data} \\
  CXR14              & \checkmark & \checkmark & \checkmark & \checkmark & \checkmark & \checkmark &\checkmark &\checkmark &\checkmark & \checkmark & \checkmark & \checkmark & \checkmark \\
  IRMA               & \checkmark & -          & -          & -          &  -         & -          & -          & \checkmark & -      & -        & \checkmark & -          & -\\
  MURA               & -          & -          & -          & \checkmark &  -         & -          & \checkmark & -      & -          & -        & -          & -          & -\\
  Bone Age           & -          & \checkmark & \checkmark & -          &  -         & \checkmark & -          & -      & -          & -        & -          & \checkmark & -\\
  ImageNet           & \checkmark &  -         & -          & -          &  -         & \checkmark & -          & -      & \checkmark & -        & \checkmark & \checkmark & \checkmark\\
\midrule
OOD Train Samples    & 3088       & 8179       & 3088       & 3088       & 0          & 3088       & 35366      &  3088  & 3088       & 0        & 220906 & 225997 & 217818\\

\midrule
Atelectasis          &  81.9     &  81.7  &  81.9  &  81.8  &  81.8  &  81.9  &  81.2  &  \textbf{82.0}  &  81.6  &  80.6  &  81.1  &  80.4  &  79.8  \\
Cardiomegaly         &  90.3     &  90.3  &  90.6  &  89.4  &  90.1  &  90.1  &  90.7  &  90.8  &  \textbf{91.0}  &  88.7  &  \textbf{91.0}  &  89.9  &  89.2  \\
Consolidation        &  80.5     &  80.1  &  \textbf{81.0}  &  80.8  &  80.3  &  80.7  &  79.4  &  80.6  &  79.8  &  79.5  &  79.7  &  80.3  &  78.9  \\
Edema                &  \textbf{89.3}  &  88.9  &  88.5  &  89.1  &  88.8  &  88.5  &  88.0  &  88.3  &  88.7  &  88.1  &  86.3  &  87.0  &  86.5  \\
Effusion             &  88.0     &  87.9  &  \textbf{88.1}  &  87.7  &  88.0  &  88.1  &  87.7  &  88.0  &  87.5  &  87.8  &  87.1  &  87.5  &  87.3  \\
Emphysema            &  91.4     &  \textbf{92.2}  &  91.4  &  92.0  &  91.6  &  91.3  &  91.1  &  90.2  &  91.2  &  87.0  &  87.8  &  87.3  &  87.4  \\
Fibrosis             &  \textbf{82.6}  &  82.1  &  79.9  &  82.1  &  82.1  &  80.4  &  82.3  &  79.6  &  79.3  &  79.0  &  79.4  &  77.9  &  78.3  \\
Hernia               &  93.6  &  \textbf{94.3}  &  91.5  &  92.8  &  91.3  &  90.9  &  92.9  &  89.7  &  91.1  &  84.2  &  82.6  &  83.5  &  82.1  \\
Infiltration         &  71.0  &  \textbf{71.3}  &  71.1  &  70.8  &  70.2  &  71.1  &  70.3  &  70.8  &  69.2  &  69.1  &  70.1  &  69.3  &  69.0  \\
Mass                 &  83.7  &  84.7  &  \textbf{84.8}  &  84.4  &  84.2  &  83.9  &  84.5  &  84.5  &  84.0  &  82.6  &  82.7  &  81.7  &  80.6  \\
Nodule               &  77.4  &  77.9  &  \textbf{78.5}  &  75.7  &  77.3  &  77.6  &  76.9  &  77.9  &  76.6  &  75.0  &  74.7  &  73.8  &  72.8  \\
Pleural Thickening   &  77.6  &  77.1  &  77.7  &  77.6  &  77.3  &  77.5  &  78.1  &  \textbf{78.2}  &  75.7  &  76.4  &  75.6  &  75.0  &  74.8  \\
Pneumonia            &  \textbf{77.7}  &  75.9  &  77.6  &  75.5  &  76.0  &  76.8  &  75.0  &  76.4  &  76.1  &  75.6  &  74.8  &  74.3  &  74.2  \\
Pneumothorax         &  86.9  &  86.6  &  86.2  &  87.3  &  \textbf{87.4}  &  86.7  &  85.6  &  86.2  &  86.4  &  85.1  &  84.7  &  85.3  &  84.5  \\
No  Finding          &  77.6  &  78.2  &  \textbf{78.3}  &  77.9  &  78.1  &  78.0  &  77.9  &  78.0  &  77.7  &  77.7  &  77.8  &  77.4  &  77.3  \\
\midrule
\textbf{Mean  AUC}            &  \textbf{83.3}  &  \textbf{83.3}  &  83.2  &  83.0  &  83.0  &  82.9  &  82.8  &  82.7  &  82.4  &  81.1  &  81.0  &  80.7  &  80.2  \\
\bottomrule
\end{tabular}
\caption{
  CXR14 classification performance evaluated by the AUC.
  Experiments are sorted by mean AUC, with the best AUC highlighted in \textbf{bold}.
  The CheXnet baseline method achieved a mean AUC of 83\%.
  Our proposed in-distribution voting (IDV) method, which trained with few (3088) OOD images, had no clear negative effect on CXR classification, with mean AUCs ranging from 82.4\% to 83.3\%.
  However, training with an OOD data set larger than the in-distribution data set reduced the mean classification AUC by up to three percentage points.
  Additionally, incorporating self-supervised heads (SS OOD) had a negative effect on the classification AUC by two percentage points.
    \\
  AUC = area under the ROC curve,
  OOD = out-of-distribution,
  CXR = Chest X-ray.
  }\label{tab:all_cxr14_aucs}
\end{table*}
\end{landscape}

We trained the CheXnet model successfully on the Chest X-ray 14 (\textbf{CXR14}) data set and evaluated the impact of training with OOD data on chest disease classification performance by measuring the performance on the ID test data set, without any OOD samples.
Furthermore, we report the CXR classification results when training with SS OOD heads.

We report the mean AUC over all 15 classes, as well as the AUC for each individual class.
The model achieved a mean AUC of 83\% when trained and tested on the CXR14 data set without any OOD images, as shown in Table~\ref{tab:all_cxr14_aucs}.
Figure~\ref{fig:roc_cxr14} displays the corresponding receiver operating characteristic (ROC) curves for all 15 classes in the CXR14 data set.

To further evaluate the performance of the model, we trained it with additional self-supervised heads, as it modifies the training procedure, which resulted in a reduction in the classification mean AUC to 81.1 \%.
We also tested the model's performance when trained with OOD samples from the IRMA and ImageNet data sets.
In contrast to the self-supervised training scheme, the mean AUC improved to 83.3\% when including IRMA and ImageNet OOD data (3088 samples).
Trained with only IRMA or ImageNet data the mean AUC resulted in 82.7\% and 82.4\%, respectively.
Table~\ref{tab:all_cxr14_aucs} shows the AUCs obtained when training with all data set combinations, including MURA and Bone Age, which resulted in similar AUCs of 83.2\% and 83.0\%, respectively.
However, using the full ImageNet OOD set resulted in a lower AUC of 81.1\%.

\begin{figure*}[h!]
\centering
\includegraphics[width=\textwidth]{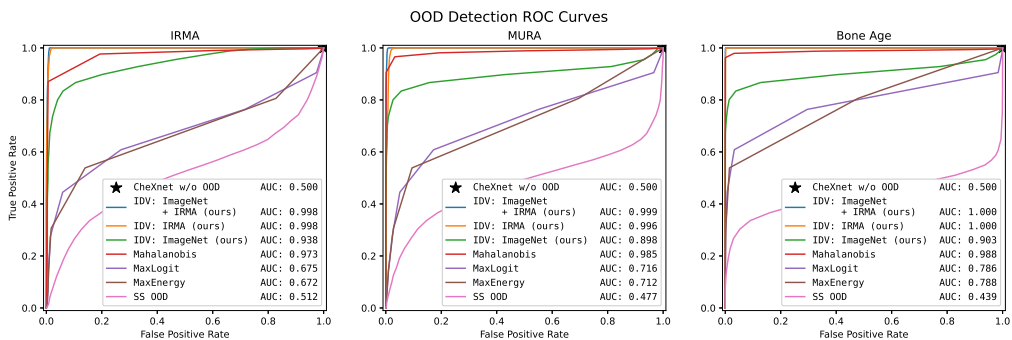}
\caption{
  ROC curves for OOD detection on the test datasets of CXR14 + {IRMA, MURA, Bone Age} with their respective AUC.
  The CXR classifier, CheXnet, cannot handle OOD data itself, resulting in a false positive rate of 100\% on all test datasets. This means that all OOD images were classified as having lung disease by the base model. Training the model with self-supervised heads (SS OOD) improved the OOD detection AUC only on the IRMA dataset, not on MURA and Bone Age. Converting the model's output to an OOD detection score (MaxLogit, MaxEnergy) improved the OOD AUC on all three datasets. Using the Mahalanobis distance to the class means in the feature space as an OOD signal resulted in an AUC greater than 97\% on all three datasets. Our proposed method, IDV, performed best with an average OOD detection AUC of 99.9\% across all three datasets when trained with ImageNet and IRMA data. Training with a domain-specific OOD dataset (IRMA) performed better than using only a general dataset (ImageNet), and training with a diverse OOD dataset containing domain-specific OOD data as well (ImageNet + IRMA) performed best. All IDV runs were trained with a subset (3088 images) of the available OOD training data, using 1044 ImageNet and 1044 IRMA images in the case of ImageNet + IRMA.
  \\
  ROC = Receiver Operating Characteristic,
  OOD = out-of-distribution,
  AUC = area under the ROC curve,
  CXR = Chest X-ray,
  IDV = in-distribution voting,
  CXR14 = Chest X-Ray 14,
  IRMA = image retrieval in medical applications data set,
  MURA = musculoskeletal radiographs data set.
}\label{fig:ood_roc}
\end{figure*}

\subsection{Out-of-Distribution Detection}

The objective of OOD detection is to classify each image as either ID or OOD.
For each of the three OOD data sets (IRMA, MURA, and BoneAge), we evaluate the performance of the OOD detection methods by measuring how many ID and OOD samples from the test set are correctly classified as such.
As a baseline, we employed the default CheXnet model with no extra OOD detection mechanism, which represents the current CXR classification models.
We report the AUC as our evaluation metric.

Figure~\ref{fig:ood_roc} shows the ROC plots for the three different OOD data sets with their corresponding AUCs.
CheXnet, without any OOD detection method, failed to filter any OOD image in all data sets, with a false positive rate of 100 \% and an AUC of 50 \%.

Training CheXnet with self-supervised heads (SS OOD) increased the OOD AUC to 51.2 \% on the IRMA data set but resulted in a worse OOD detection performance on the MURA and Bone Age data sets with 47.7 \% AUC and 43.9 \% AUC, respectively.

The conversion of the logits to an OOD score using MaxLogit and MaxEnergy improved the OOD performance considerably compared to the CheXnet model with MaxLogit achieving AUCs of 67.5 \%, 71.6 \%, and 78.6 \%, respectively on the IRMA, MURA, and Bone Age data sets. MaxEnergy performed similarly, with 67.2 \%, 71.2 \%, and 78.8 \%, respectively.

Using the Mahalanobis distance increased the OOD detection performance significantly compared to MaxLogit and MaxEnergy with an AUC of 97.3 \% on the IRMA data set, 98.5 \% on MURA, and 98.8 \% on Bone Age.

Our method, IDV, trained with 1544 ImageNet and 1544 IRMA images surpassed the Mahalanobis performance on all three data sets with an AUC of 99.8 \% on the IRMA data set, 99.9 \% on the MURA data set, and 100 \% on the Bone Age data set.
Training with IRMA images resulted in 99.8 \%, 99.6 \%, and 100 \% AUC on IRMA, MURA, and Bone Age, respectively.
IDV with ImageNet in 93.8 \%, 89.8 \%, and 90.3 \%, respectively.

\subsection{Effect of Out-of-Distribution Training Data}

To investigate the impact of out-of-distribution training data selection, we conducted a series of experiments to measure the performance of OOD detection, using the area under the receiver operating characteristic curves.
We trained our model using all available OOD data sets, including IRMA, MURA, Bone Age, and ImageNet.
To account for the smaller sample size of IRMA and Bone Age data sets, we combined them with the ImageNet data set.
Specifically, we chose ImageNet training data larger than the in-distribution CXR14 training data to measure the effect of using more OOD than ID data.
To ensure a fair comparison among different OOD training data sets, we randomly sampled a subset of 3088 images from each data set, matching the smallest data set size (IRMA).
When training using two data sets (ImageNet + IRMA, ImageNet + Bone Age), we selected 50\% from each, resulting in 1544 images from ImageNet and 1544 radiographs from either IRMA or Bone Age.
In all experiments, we used a test set consisting of 3860 samples (see Table~\ref{table:data set}).

\begin{figure*}
\centering
\includegraphics[width=\textwidth]{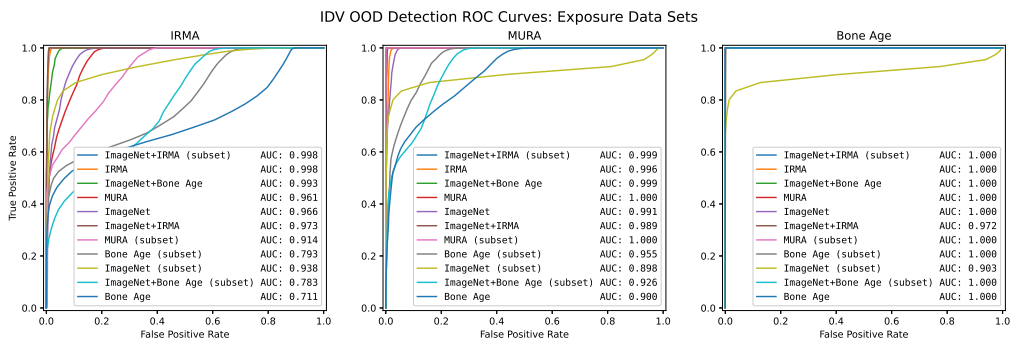}
\caption{
  ROC curves and AUCs of all OOD detection runs using IDV on three OOD test data sets: IRMA, MURA, and Bone Age with CXR14 in-distribution data.
  IDV OOD detection with any OOD data improved OOD detection performance.
  Generally, all models performed best on the Bone Age data set, which includes only hand X-rays, and the worst on IRMA, which comprises a variety of X-rays.
  Consequently, using only the specific Bone Age data during training improved OOD detection performance less than using the diverse ImageNet data set, expect on the Bone Age test data.
  Training with ImageNet OOD images provided strong OOD detection performance, with an AUC greater than 96 \% on all data sets.
  Additionally, training with the most diverse data set, ImageNet + IRMA, provided the overall best performance, using only 3088 training images (subset).
    \\
  ROC = Receiver Operating Characteristic,
  OOD = out-of-distribution,
  AUC = area under the ROC curve,
  CXR = Chest X-ray,
  IDV = in-distribution voting,
  IRMA = image retrieval in medical applications data set,
  MURA = musculoskeletal radiographs data set.
}\label{fig:idv_ood_roc}
\end{figure*}

Figure~\ref{fig:idv_ood_roc} shows the ROC curves and AUC values for the different IDV runs evaluated on the thee test data sets: IRMA, MURA, and Bone Age.
Generally, our proposed method outperformed the CheXnet baseline (AUC 50 \%) when trained on any OOD data.
Also, the models performed best on the Bone Age data set, containing only hand X-rays, and worst on the IRMA data set, containing a wide variety of radiographs.
Training with ImageNet data generalized to X-ray OOD data sets with AUCs from 97 \% to 100 \% when trained on the whole data set and from 90 \% to 94 \% when trained on the subset.
Training on the Bone Age data set performed worst on the IRMA data set (AUC 71 \%) but achieved an AUC of 100 \% on the Bone Age test set; using only a subset improved the performance and combining the data with ImageNet even further.
Training on the MURA and IRMA data set individually performed best, but was exceeded only by the combination of ImageNet and IRMA data.
Overall, our results suggest that incorporating diverse data sets, such as ImageNet and IRMA, is a promising approach to improve OOD generalization for X-ray classification tasks.

\section{Discussion}
Assessing whether the tested model performance in a benchmark translates to an intended production setting, including potential OOD data is a necessary step before deploying a machine learning model.
This is particularly important in safety critical applications, e.g. when classifying chest X-rays to assist radiologists in diagnosing patients.
Our results show that the CheXnet model cannot handle OOD samples out-of-the-box.
However, combining it with our proposed method, IDV, and trained with any OOD data, even the photographs of ImageNet, improved the OOD detection performance compared to the baseline CheXnet model and most OOD detection methods considerably with OOD detection AUCs up to 100~\%.

In their paper, Rajpurkar et al. conclude that their CheXnet model exceeds practicing radiologists in detecting pneumonia~\cite{rajpurkarChexnetRadiologistlevelPneumonia2017} and note as limitation that only frontal CXR were used, giving a potential low estimate of the model's performance.
Our experiments showed that a further limitation was not considered: out-of-distribution images.
A model that cannot handle OOD images, making confident predictions based on wrong evidence, will lead to worse quality of care, eroding the trust of physicians into the model's predictions when facing ID images, and impede the potential benefits of computer assisted diagnosis.
Having robust models trusted by radiologists is necessary to leverage such classifiers to assist radiologists in clinical practice.
Out-of-the-box, the CheXnet model failed to provide this robustness against realistic OOD images.

In our experiments, we noted that other OOD detection methods based on a model's output (Maxlogit, MaxEnergy, and SS OOD) performed considerably worse than in their original works.
This suggests that the presented OOD data sets are more challenging, highlighting the importance of considering OOD data in the medical domain.
Our experiments showed that even limited OOD data leveraged by IDV had a large effect on the OOD detection performance without negatively affecting the intended classification task.

We interpret the large OOD detection difference between the output based methods, MaxLogit and MaxEnergy, and Mahalanobis as evidence for the ``label overfitting'' hypothesis.
Our approach, including an ``no finding'' / OOD label into the training procedure, breaks this overfitting problem and improves the OOD detection performance without a complex clustering component like the Mahalanobis distance.
We interpret the IDV results as indicating that the model incorporates the fact that OOD images exist into its output.

While training with OOD data improves OOD detection performance, it is important to consider the intended use-case: CXR classification.
We can conclude that training with OOD data, as proposed in our method IDV, does not affect chest disease classification performance negatively.
This means that diversifying the training and validation data set with OOD samples improves the model performance in real-world scenarios, as potential OOD images are filtered.

Regarding the choice and availability of OOD training data our results showed that only few thousand samples are sufficient and even unrelated OOD data, such as ImageNet, is immensely useful.
When comparing the OOD performance trained on the very specific Bone Age data set and the general ImageNet for cross-data set generalization we note that training with unrelated ImageNet data generalized better.
We therefore conclude that using a generic OOD data set alone could improve a model's OOD detection performance.
Including domain specific, diverse OOD images improves the OOD detection AUC even further (cf. ImageNet vs. ImageNet + IRMA in Figure~\ref{fig:idv_ood_roc}).

In this work, we investigated the effect of OOD images on a chest X-ray classifier.
We showed that the model, reportedly performing as good as radiologists~\cite{rajpurkarChexnetRadiologistlevelPneumonia2017, rajpurkarDeepLearningChest2018}, was not able to filter OOD images, leading to obvious false positives to the human observer.
We assume its predictions are conditioned on chest X-rays, because the model was only trained on chest X-rays, leading to overconfident predictions given OOD images.
As hypothesized by Lee et al.~\cite{leeSimpleUnifiedFramework2018}, this leads to an ID-overfitted output space.
This interpretation explains why established output-based OOD detection methods failed in our experiments, when compared to detecting OOD samples in the feature space.
Our solution, ID voting and training with OOD images, regularizes the output space and expands the model's knowledge horizon, leading up to a 100 \% ID OOD detection AUC.

One reason why OOD data are rarely considered is their dependency on the intended application.
We showed that including a small OOD training data set from the same data set as the OOD test data resulted in a better OOD detection performance than a general OOD data set.
While this suggests that there is no ideal application independent OOD data set, we found that training with any OOD data improved the baseline performance considerably.
Furthermore, we showed that even a few thousand OOD samples from the intended application boosted the specificity considerably.
Therefore, when creating a data set to train and evaluate a model in a production setting, we recommend to remove anomalies, outliers and other OOD with caution.
Instead, including this ``real-world'' data not only in the training process, but also into the model validation, will lead to more robust ML models and ultimately improve clinical acceptance

One limitation of this work is that we use the CheXnet model as a representative for other chest X-ray classification models.
While we argue that this architecture is a strong baseline, further research is necessary to determine if our findings translate to other architectures.
Furthermore, we only tested our model on chest X-ray images, even though our approach remains relevant to all multi-label OOD detection data sets.
Finally, this retrospective work was performed only on public data and further work is necessary to evaluate our findings on real-world clinical data.

\section{Conclusion}
In conclusion, our study demonstrated that training solely on ID data leads to incorrect classification of OOD images as ID, resulting in increased false positive rates.
We also showed, that our proposed method, IDV, substantially improves the model's ID classification performance, even when trained with data that will not occur in the intended use case or test set.
Consequently, our approach makes the final model more robust and considerably improves its predictive performance in a real-world setting.

\section{Acknowledgments}
This work was supported in part by the German federal ministry of health’s program for digital innovations for the improvement of patient-centered care in healthcare [grant agreement no. 2520DAT920].

\bibliographystyle{./medphy.bst}
\end{document}